\documentclass[11pt,twoside]{article}  

\usepackage{adassconf}

\usepackage[dvips]{graphicx}

\usepackage{array}

\begin{document}




\paperID{P7-9}




\title{Age determination of the nuclear stellar population of Active Galactic Nuclei using Locally Weighted Regression}

\titlemark{Age determination of the nuclear stellar population of AGN using LWR}




\author{Trilce Estrada-Piedra, Juan Pablo Torres-Papaqui, Roberto Terlevich, Olac Fuentes, Elena Terlevich}

\affil{Instituto Nacional de Astrof\'\i sica \'Optica y Electr\'onica, Puebla, 72840, Mexico}









%



%

\contact{Trilce Estrada-Piedra}

\email{trilce@inaoep.mx}




\paindex{Estrada-Piedra, T.}

\aindex{Torres-Papaqui, J. P.}

\aindex{Terlevich, R.}

\aindex{Fuentes, O.}

\aindex{Terlevich, E.}






%






%




\authormark{Estrada-Piedra, Torres-Papaqui, Terlevich, Fuentes, Terlevich}







\keywords{LWR, SFBS, clasification, astronomy: age determination, AGN, nuclear stellar population }




\begin{abstract}          

We present a new technique to segregate old and young  stellar populations in galactic spectra using machine learning methods. We used an ensemble of classifiers, each classifier in the ensemble specializes in young or old populations and was  trained with locally weighted regression and tested using ten-fold cross-validation. Since the relevant information concentrates in certain regions of the spectra we used the method of sequential floating backward selection offline for feature selection.

The application to Seyfert galaxies proved that this technique is very insensitive to the dilution by the Active Galactic Nucleus (AGN) continuum. Comparing with exhaustive search we concluded that both methods are similar in terms of accuracy but the machine learning method is faster by about two orders of magnitude.

\end{abstract}




\section{Introduction}

Recent spectroscopic surveys of nearby AGN have proven that a large fraction show high-order hydrogen Balmer absorption lines in the near-UV (Gonz\'alez-Delgado et al 1999) (Joguet et al 2001). These features are characteristic of young stars and therefore represent strong evidence of recent star formation in these galaxies. 

From a theoretical point of view, it is very important to determine the age of these starbursts, in order to understand the nature of the starburst-AGN connection and galaxy formation and evolution. The characterization of the nuclear star forming region (its age and mass) is very difficult to achieve in AGN, due to the contamination of the nuclear stellar absorption lines by the AGN component itself. The recent release of high-resolution spectra of large number of galaxies by the Sloan Digital Sky Survey (SDSS) consortium allows spectroscopic studies to be performed now on thousands of galaxies with  active nuclei.

In this work we propose a Machine Learning (ML) method to determine the age of stellar populations in synthetic galactic spectra. The experimental results obtained here show the efficiency of the automatic learning method applied to astronomy.

\section{Background}

\subsection{Sequential Floating Backward Selection (SFBS)} \label{sfs}

SFBS is a feature selection algorithm that allows to work with non-monotonic data. It constructs in parallel the feature sets of all dimensionalities up to a specified threshold and consists of applying after each feature exclusion a number of features inclusion as long as the resulting subsets are better than those previously evaluated at that level. It makes a dynamically controlled number of iterations and achieves good results without static parameters (Pudil 1994).

\subsection{Locally Weighted Regression (LWR)} \label{lwr}

LWR is an instance based learning method; it assumes instances can be represented as points in an Euclidean space (Moore 2001). Its training consists of explicitly retaining the training data and using them each time a prediction needs to be made. LWR performs a regression around a point of interest using only a local region around that point. Locally weighted regression can fit complex functions in an accurate way and data modifications have little impact on the training.

\subsection{Ensembles of Classifiers}\label{ensembles}

An ensemble of classifiers is a group of classifiers trained independently whose outputs are combined in some way, usually by voting (Mitchell 1997). They are normally more accurate than the individual classifiers that make it up.

\section{Data}

The data is composed by 14 high resolution synthetic spectra combined in pairs considering 10 levels of dilution and including Gaussian noise. The experiments were calibrated using two population synthesis models with different ages (A.~Bressan, private communication):

\begin{itemize}

\item A young population with ages between $10^{7.0}$ and $10^{8.6}$ years, representing a starburst component,

\item An old population with ages between $10^{8.0}$ and $10^{9.6}$ years, representing the bulge component.

\end{itemize}

The spectra consist of $5655$ points, from the optical to the near UV with wavelengths between $3600$ and $5300$\AA\ and $0.2\AA$  sampling. A subsampling process was performed to a resolution of $1\AA$ in order to make the data compatible with the resolution of the observed spectra, in our case the SDSS spectral data. The features selected are the Balmer and Calcium II lines that are characteristic of the young and old populations respectively. The points of interest are selected for each component and population using the SFBS algorithm.

\section{Implementation}

\begin{figure}[!htb]

\begin{center}

\includegraphics[width=9.5cm,height=5cm]{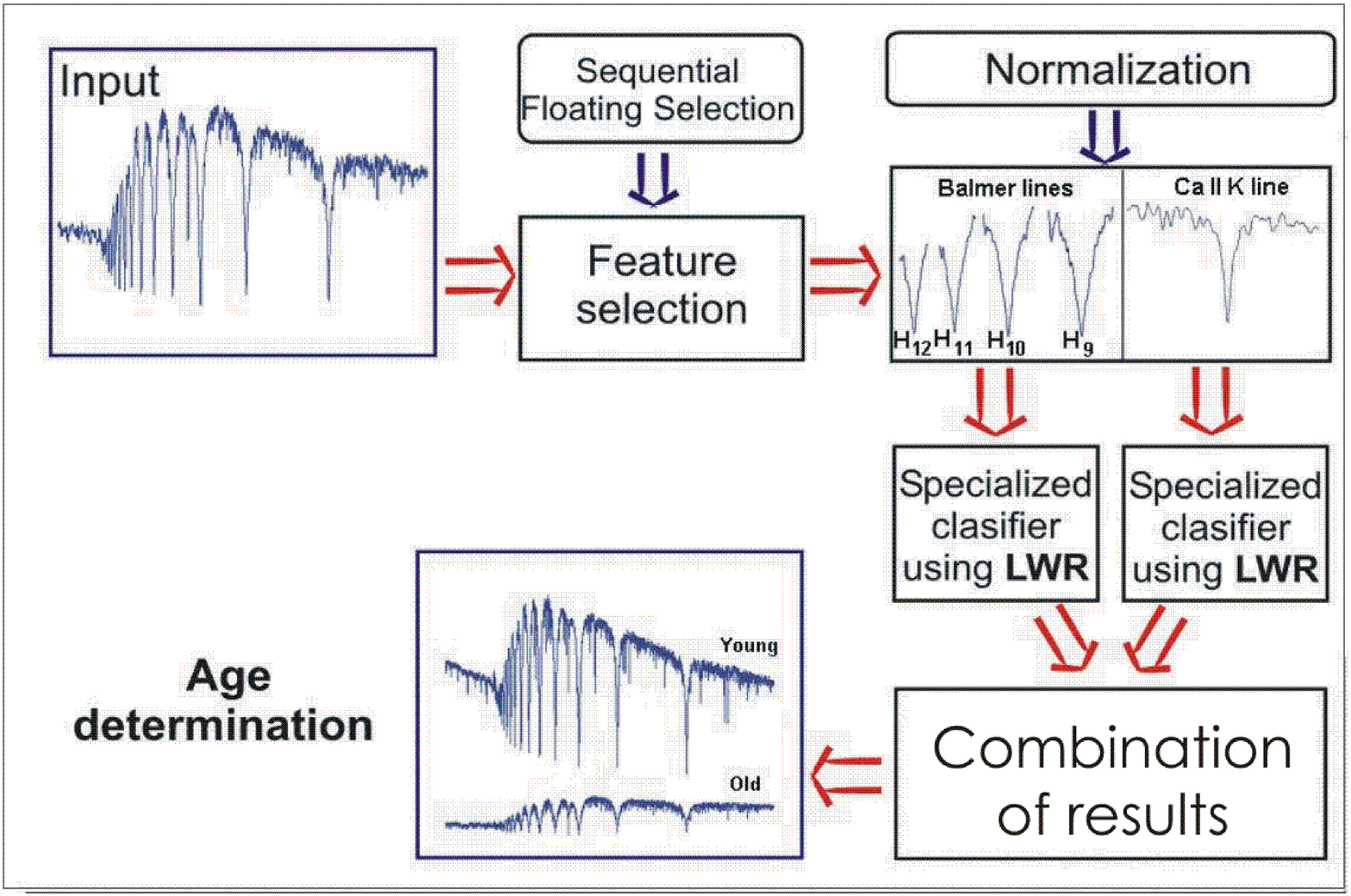}

\caption{Process of age identification}\label{fprincipal}

\end{center}

\end{figure}

We made specialized classifiers using LWR trained in the high-order Balmer and Ca II absorption lines to determine the young and the old populations respectively; at the end of the process the results were combined by an ensemble of classifiers. For each classifier we selected a subset of features maximizing the probability of correct classification, this goal was achieved applying the SFBS method offline. The general process of age identification is shown in Figure \ref{fprincipal} and is as follows:

 \begin{enumerate}

\item Select features in the relevant regions. This is done online using the information retrieved by SFBS in a previous step.

\item Extract the information in the Ca II K line, which is typical of the older bulge population.

\item Identify the old population using specialized classifiers.

\item For each spectrum with no classification: extract the information in the Balmer lines that is characteristic of recent star formation.

\item Identify the young population using specialized classifiers.

\item Combine results in an ensemble.

\end{enumerate}

\section{Experimental Results}

\begin{table}[!htb]

\caption[]{Summary of results.}\label{tresults}

\begin{center}

\begin{tabular*}{\textwidth}{
@{}c
@{\extracolsep{\fill}}r
@{\extracolsep{\fill}}c
@{\extracolsep{\fill}}c
@{\extracolsep{\fill}}c
@{\extracolsep{\fill}}c
@{\extracolsep{\fill}}c@{}}
\tableline\tableline\vrule width 0pt height 2.5ex depth 1ex
           &       & \multicolumn{2}{c}{Young} & \multicolumn{2}{c}{Old} & Time \\
Algorithm  & Noise &  Sample & True &  Sample & True & sec. \\
           &       & Error & Error & Error & Error &         \\
\hline
Exhaustive Search & 0-11  &  -  &  0 &  - & 0 & 12 \\
\hline
Machine Learning  &  0 & 0.005 & [0.003,0.01] & 0.001 & [0.003,0.01] & 0.014 \\
     $0.2\AA$     &  3-11\% & 0.007  & [0.001,0.01] & 0.025 & [0.01,0.04] & 0.092\\
\hline
Machine Learning  &  0 & 0.039  & [0.02,0.05] & 0.049 & [0.03,0.06] & 0.16\\
       $1\AA$     & 3-11\% & 0.059 & [0.04,0.077] & 0.069 & [0.05,0.08] & 0.35 \\
\hline
\end{tabular*}
\end{center}
\end{table}

We experimented using data with different resolutions and adding different noise levels. First we used $14$ high spectral resolution synthetic models at $0.2\AA$, after that we sub-sampled the spectra to $1\AA$ to evaluate the performance of the ML method and to decide if it is possible to make an extension to handle the SDSS spectra. 

The ensemble was tested using ten fold cross-validation; in ten fold cross validation we divide the data into 10 subsets of equal size; we train the classifier 10 times, each time leaving out one of the subsets and using it for testing the algorithm. The sample error is calculated each time and is averaged to obtain the true error. An accuracy of about $0.3~dex$ in logarithmic age was achieved. The main results are summarized in Table \ref{tresults}.  

The time of prediction using LWR is linear in the number of examples and the experiments show that the time is reduced drastically with respect to the technique that does not involve ML. The method was then applied to the optical/near UV spectra of nuclear regions of nearby Seyfert galaxies covering the wavelength region $3600-5300\AA$ and it was found to be rather insensitive to the emission line and continuum contamination.

\section{Conclusions and Future Work}

The results obtained by ML are compared with those produced by exhaustive search in terms of time and precision. The machine learning method could find the correct ages, and was faster than exhaustive search, and has the additional advantage of the generalization capabilities inherent to this kind of algorithms. We conclude that the ML method can be extended to work with real spectra if we use a realistic noise model. The two-dimensional classification used for age identification in galactic nuclear spectra is similar in many ways to other problems and can be taken as a guideline in different problems (for example classification of binary stars and search for supernovae in galactic spectra).

An extension of the method to handle observational spectra more reliably is to be published elsewhere. A second goal will be to implement a ML algorithm to find the AGN dilution and modify the synthetic spectra to predict three ages of stellar populations instead of just two.

\acknowledgments This work was partially supported by CONACyT (the Mexican Research Council) under grants 31877, 171595 and 32186-E.
E-P and T-P gratefully acknowledge financial support from the Conference organizers.





\end{document}